\documentclass[twocolumn,showpacs,preprintnumbers,amsmath,amssymb]{revtex4}

\usepackage{dcolumn}
\usepackage{bm}

\usepackage{amsmath}
\usepackage{amsfonts}   
\usepackage{graphics}  
\usepackage{psfrag} 
\usepackage{graphicx} 
\usepackage{epsfig}    
\usepackage{rotating}  
 \usepackage[latin1]{inputenc}
 \usepackage{color}
 \usepackage{rotating}

\usepackage{color}
\usepackage{epsfig}
\definecolor{darkgreen}{rgb}{0,0.5,0} 
\definecolor{violet}{rgb}{0.5,0,0.5}
\definecolor{orange}{rgb}{0.2,0.5,0.5}

\begin{document}

\preprint{}

\title{Evolutionary game theory in growing populations
}

\author{Anna Melbinger, Jonas Cremer, and Erwin Frey}
\affiliation{Arnold Sommerfeld Center for Theoretical Physics (ASC) and Center for NanoScience (CeNS), Department of Physics, Ludwig-Maximilians-Universit\"at M\"unchen, Theresienstrasse 37, D-80333 M\"unchen, Germany}

\date{\today}
\begin{abstract}
Existing theoretical models of evolution focus on the relative fitness advantages of different mutants in a population while the dynamic behavior of the population size is mostly left unconsidered.
We here present a generic stochastic model which combines the growth dynamics of the population and its internal evolution. Our model thereby accounts for the fact that both evolutionary and growth dynamics are based on individual reproduction events and hence are highly coupled  and stochastic in nature.
We exemplify our approach by studying the dilemma of cooperation in growing populations and show that genuinely stochastic events can ease the dilemma by leading to a transient but robust increase in cooperation. 
\end{abstract}

\pacs{87.23.Kg, 87.10.Mn, 05.40.-a}
\maketitle

Commonly, Darwinian evolution in terms of reproduction, selection, and variation is described in frameworks of population genetics and evolutionary game theory~\cite{Ewens,Maynard,Blythe:2007}.
These approaches model the  internal evolutionary dynamics of a species' different strategies (or traits) in a relative perspective. 
Namely, 
they compare fitness terms and focus on the relative advantage and abundance of different traits. In such a setup, the time evolution of the relative abundance, $x$, of a certain strategy is frequently described by a \emph{replicator equation},
\begin{equation}
\partial_t x=\left(f-\langle f \rangle\right)x.
\label{eq:replicator}
\end{equation}
A trait's relative abundance will increase if its fitness $f$ exceeds the average fitness $\langle f \rangle$ in the population. 

While in these evolutionary approaches,  the dynamics of the population size, $N$, is mostly left unconsidered or assumed to be fixed~\cite{Blythe:2007}, in population ecology the dynamical behavior of a species' population size is studied.
Models of population dynamics~\cite{Murray, Hastings1997} usually describe the time development of the total number of individuals, $N$, by equations of the form 
 \begin{equation}
\partial_t N=\mathcal F\left(N,t\right).
\label{eq:pop_dyn}
\end{equation}
$\mathcal F\left(N,t\right)$ is in general a non-linear function which includes the influence of the environment on the population, such as the impact of restricted resources or the presence of other species. By  explicitly depending on time a changing environment such as, for example, the seasonal variation of resources can be taken into account.

The internal evolution of  different traits and the dynamics of a species' population size are, however, not independent~\cite{Hibbing2010}. Actually, species typical coevolve with other species in a changing environment and a separate description of both evolutionary and population dynamics is in general not appropriate. Not only population dynamics affects the internal evolution (as considered, for example, by models of density-dependent selection~\cite{Roughgarden}), but also vice versa. Illustrative examples of the coupling are biofilms which permanently grow and shrink. In these microbial structures diverse strains live, interact, and outcompete each other while simultaneously affecting the population size~\cite{West:2006p1249}. So far, specific examples of this coupling have been considered by deterministic approaches only, e.g.~\cite{Hauert3,Cressman}. However, classical and recent work has emphasized the importance of fluctuations for internal evolution~which are only accounted for by stochastic, individual-based models, e.g.~\cite{Kimura,nowak-2004-428,Traulsen,Cremer2}. 

In this Letter, we introduce a class of stochastic models which consider the interplay between population growth and its internal dynamics. 
Both processes are based on reproduction events. A proper combined description should therefore be solely based on \emph{isolated} birth and death events. Such an approach also offers a more biological interpretation of evolutionary dynamics than common formulations like the Fisher-Wright or Moran process~\cite{Ewens,Blythe:2007,Moran,nowak-2004-428}. That is to say, fitter individuals prevail due to higher birth rates and not by winning a tooth-and-claw struggle where the birth of one individual directly results in the death of another one. The advantage of our formulation is illustrated by  the dilemma of cooperation where a transient increase in cooperation can be found (which does not exist in standard approaches,  Eq.~(\ref{eq:replicator})).

In the following, we consider two different traits, $A$ and $B$, in a well-mixed population, all the same generalizing the model to more traits is straightforward. The state of the population is then described by the total number of individuals $N=N_A+N_B$ and the fraction of one trait within the population, $x=N_A/N$. The stochastic evolutionary dynamics is fully specified by stochastic birth and death events with rates
\begin{eqnarray}
\Gamma_{\emptyset \to S}=G_S(x,N) N_{S},\hspace{0.5cm} \Gamma_{S\to\emptyset}&=&D_S(x,N) N_{S} \, ,
\label{eq:rates}
\end{eqnarray}
where $G_S(x,N)$ and $D_S(x,N)$ are per capita reproduction and death rates for an individual of type $S\in\{A,B\}$, respectively. We consider these rates to be separable into a global and relative part, meaning a 
trait-independent and trait-dependent part: 
\begin{align}
G_S=g(x,N)f_S(x),\hspace{0.3cm}D_S=d(x,N)w_S(x).
\end{align}
The \emph{global population fitness}, $g(x,N)$, and the \emph{global population weakness}, $d(x,N)$, affect the population dynamics of all traits in the \emph{same} manner. For example, they account for constraints imposed by limited resources or how one strategy impacts the whole population. In contrast, the \emph{relative fitness}, $f_S(x)$, and the \emph{relative weakness},  $w_S(x)$, characterize the relative advantage of one strategy compared to the other. They are different for each trait and depend, in a first approach, only on the relative abundance $x$ \footnote{By incorporating a $N$-dependence in $f_S$ and $w_S$, one can extend our model to other forms of density-dependence.}. The relative fitness terms, $f_S(x)$, affect the corresponding birth rates, and the relative weakness functions,  $w_S(x)$, describe the chances for survival of distinct traits.

While in evolutionary game theory  only the relative fitness is considered~\cite{Maynard}, and common models of population dynamics take only the global functions into account, we here consider both global and relative fitness and show how their interplay determines the evolutionary outcome of a system. In the following, we set  $w_A(x)=w_B(x)=1$ in order to compare our unifying approach with standard formulations~\cite{Maynard}. Though the full stochastic dynamics are given by a master equation, it is instructive to disregard fluctuations for now  and examine the corresponding set of deterministic rate equations:
\begin{subequations}
\begin{align}
\partial_t x &= g(x,N)\left(f_A(x)-\left<f \right> \right)x,\label{eq:mfx}\\
\partial_t N &= \left[g(x,N)\langle f\rangle-d(x,N)\right]N\label{eq:mfN} ,
\end{align}
\label{meanfield}
\end{subequations}
where $\langle f\rangle=xf_A+(1-x)f_B$ denotes the average fitness. Eq.~\eqref{eq:mfx} has the form of a replicator equation~\cite{Maynard}. However, in Eq.~(\ref{eq:mfx}) there is an additional factor, namely the global population fitness $g(x,N)$. This leads to a coupling of $x$ and $N$ whose implications we will discuss later on. Similarly, Eq.~\eqref{eq:mfN} describing population growth is coupled to the internal evolution, Eq.~\eqref{eq:mfx}. Note that for frequency-independent global functions, $g(x,N)\equiv g(N)$ and $d(x,N)\equiv d(N)$, Eqs.~(\ref{meanfield}) resemble  Eqs.~(\ref{eq:replicator}) and (\ref{eq:pop_dyn}). Only then, the deterministic dynamics reduces to the common scenario \cite{Moran,nowak-2004-428,Traulsen}, where a changing population size is immaterial to the evolutionary outcome of the dynamics~\cite{Blythe:2007}. For the full stochastic dynamics the strength of fluctuations scales as $\sqrt{1/N}$ ~\cite{Blythe:2007,Kimura,Cremer2} and thereby is strongly affected by population growth.

In more realistic settings, the global fitness and weakness functions, $g(x,N)$ and $d(x,N)$,  can also depend on the relative abundance, $x$. This implies an interdependence of population growth and internal evolution. In the following, we focus on one particular but very important example: the dilemma of cooperation in a growing population. There is an ongoing debate in sociobiology regarding how cooperation within a population emerges in the first place and how it is maintained in the long run~\cite{NowakCooperation,West:2006p1249}. Microbial biofilms serve as versatile model systems~\cite{West:2006p1249,Gore:2009p1538,Chuang01092009,Griffin}. There, cooperators are producers of a common good, usually a metabolically expensive biochemical product. For example, for the proteobacteria \emph{Pseudomonas aeruginosa}, cooperators produce iron-scavenging molecules (siderophores). Released into the environment these molecules strongly support the iron uptake of each individual in the population~\cite{Griffin}. Cooperators thereby clearly increase the global fitness of the population as a whole, leading to a faster growth rate and a higher maximum population size~\cite{Griffin}. In such a setting, however, non-producers (``cheaters") have a relative advantage over cooperators as they save the cost of providing the common good, e.g. the production of siderophores. Hence, their relative fraction is expected to increase within the population implying that the global fitness of the population declines. Surprisingly, as we show in the following, a coupling between growth and internal evolution can overcome this dilemma transiently and the average level of cooperators can increase despite a disadvantage in relative fitness.

We model the internal evolutionary dynamics by the prisoner's dilemma game~\cite{Maynard,NowakCooperation}. Within this standard approach, individuals are either cooperators ($A$) or cheaters ($B$). While cooperators provide a benefit $b$ to all players at the expense of a (metabolic) cost $c<b$, a cheaters save the cost by not providing the benefit. The relative fitness of these traits is given by $f_A(x)=1+s\left[(b-c)x-c(1-x)\right]$ and $f_B(x)=1+sbx$, respectively, where the frequency-independent and dependent parts are weighted by the strength of selection $s$~\cite{nowak-2004-428}. Analyzing the prisoner's dilemma per se, defectors are always better off than cooperators because of their advantage in relative fitness, $f_A(x)<f_B(x)$~\cite{NowakCooperation}. 
In the following, we choose for specificity $b=3$ and $c=1$, however, our conclusions are independent of the exact values.

Importantly, cooperation positively affects the whole population by increasing its global fitness, e.g.~by production of a common good like siderophores. Here, we consider bounded population growth with a growth rate increasing with the cooperator fraction $x$. In detail, we choose a $x$-dependent global fitness, $g(x)=1+p x$, and a $N$-dependent global weakness, $d(x,N)=N/K$ accounting for limited resources. For $p=0$, one obtains the well-known dynamics of logistic growth~\cite{Verhulst} with a carrying capacity $K$. For $p> 0$, the carrying capacity, $K(1+px)$, depends on the fraction of cooperators. For instance, for {\it P. aeruginosa}~\cite{Griffin}, the iron uptake, and hence the birth rates, increase with a higher siderophore density and therefore with a higher fraction of cooperators.

\begin{figure}
\centering
\includegraphics[width=0.44\textwidth]{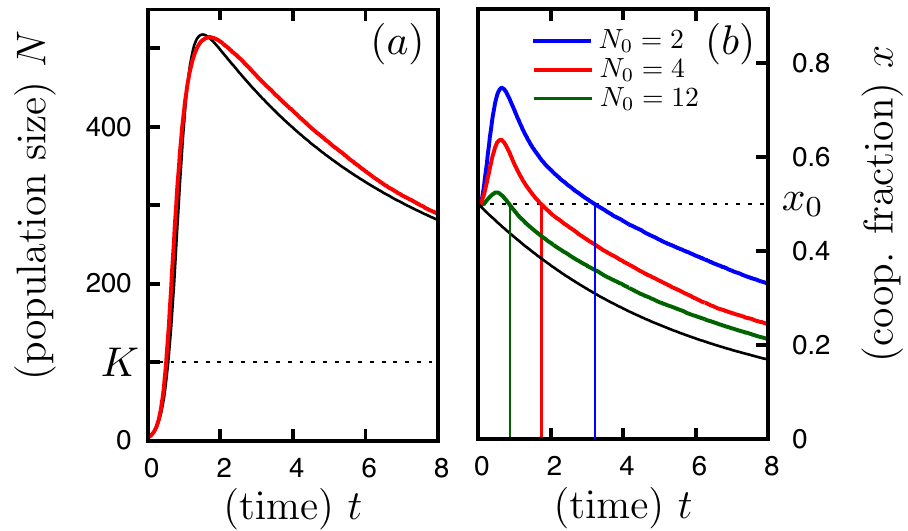}
\caption{The dilemma of cooperation in growing populations. (a) Average population size over time. Due to a cooperation-mediated growth advantage, it can show an overshoot.  The red line corresponds to simulation results while the black line is obtained by evaluating Eqs.~(\ref{eq:pd}). (b)  The average level of cooperation increases transiently for times $t<t_c$, especially if the initial population size is small meaning fluctuations are large. The parameters are given by $x_0=0.5$, $b=3$, $c=1$, $s=0.05$, $K=100$ and $p=10$. $N_0$ is $4$ (red line), $2$ (blue line), and $12$ (green line), respectively. The black line is obtained by evaluating Eqs.~(\ref{eq:pd}) for $N_0=4$. Cooperation times $t_c$ are denoted by thin lines of the corresponding color. 
\label{Fig:A}}
\end{figure}

To analyze the evolutionary behavior of our model we performed extensive simulations of the stochastic dynamics given by the master equation determined by the birth and death rates, Eqs.~(\ref{eq:rates}). All ensemble averages were performed over a set of  $10^4$ realizations. In Fig.~\ref{Fig:A} the average population size, $N$, and the average fraction of cooperators, $x$, are shown for different initial population sizes, $N_0$. The influence of a frequency-dependent growth on the population is twofold.
First, starting in the regime of exponential  growth, the frequency-dependent global fitness may cause an \emph{overshoot in the population size} [Fig.~\ref{Fig:A}(a)].
Second, and more strikingly, the selection disadvantage of cooperators can be overcome and a \emph{transient increase of cooperation} emerges, [Fig.~\ref{Fig:A}(b)]. It is maintained until a time $t_c$, which we term as the  \emph{cooperation time}.

Both phenomena rely on a subtle interplay between internal evolution, with a selection pressure towards more defectors, and population growth, with a growth rate increasing with the fraction of cooperators.  
While the overshoot in population size can already be understood on the basis of the rate equations,
\begin{subequations}
\begin{align}
\partial_t x&=-s(1+p x)x(1-x),\label{eq:pdx}\\
\partial_t N&=\left[(1+px)\langle f\rangle-N/K\right]N,\label{eq:pdN}
\end{align}
\label{eq:pd}
\end{subequations}
the transient increase of cooperation is a genuinely stochastic event as discussed in detail below.
A first impression of the antagonism between selection pressure and growth can already be obtained by examining the characteristic time scales. While the fraction of cooperators changes on a time scale $\tau_x\propto{1}/{s}$, the population size evolves on a time scale $\tau_N\propto 1$. Hence, the strength of selection, $s$, regulates the competition between population growth and internal dynamics. 
For $s\gg 1$, selection is much faster than growth dynamics. Therefore, the rapid ensuing extinction of cooperators cannot be compensated for by the growth advantage of populations with a larger fraction of cooperators. In contrast, in the limit of \emph{weak selection} ($s\ll1$), growth dynamics dominates selection and both an overshoot in the population size and a transient increase of cooperation become possible (see below). In the following we focus on this latter, more interesting, scenario of weak selection ($\tau_N<\tau_x$).

Let us first consider the overshoot in the population size [Fig.~\ref{Fig:A}(a)]. 
It is caused by a growth rate and a carrying capacity which are increasing functions of the fraction of cooperators (here we use $p=10$ as observed in microbial experiments~\cite{Chuang01092009}). For $t<\tau_x$, a small population ($N\ll K(1+px_0)$) with an initial fraction of cooperators, $x_0$, grows exponentially towards its comparatively large carrying capacity $K(1+px_0)$. During this initial time period the fraction of cooperators evolves only slowly and can be considered as constant. On a longer time scale, $t>\tau_x$, however, selection pressure drives the fraction of cooperators substantially below its initial value $x_0$, leading to a smaller carrying capacity, $K(1+px)$. Finally, cooperators go extinct and the population size decreases to $K$. This functional form of $N(t)$ is well described by the rate equations~(\ref{eq:pd}); see black line in Fig.~\ref{Fig:A}(a) . 

In contrast, the transient increase of cooperation, cf. Fig.~\ref{Fig:A}(b), cannot be understood on the basis of a simple deterministic approach, where $\partial_t x\leq0$ holds strictly (see black line in Fig.~\ref{Fig:A}(b)). It  is a genuinely \emph{stochastic effect}, which relies on the amplification of stochastic fluctuations generated during the initial phase of the dynamics where the population is still small. In more detail, for small populations , the fraction of cooperators is subject to strong fluctuations and differs significantly from one realization to another. Crucially, due to the coupling between the growth of a population and its internal composition, these fluctuations are amplified asymmetrically favoring a more cooperative population, i.e.~growth, set by the global fitness $g(x)$, is amplified by an additional cooperator while it is hampered by an additional defector. This implies that the ensemble of realizations becomes strongly skewed towards realizations with more cooperators. If this effect is strong enough the ensemble average $x(t)=\sum_{i}N_{A,i}(t)/\sum_i N_i(t)$, which describes the mean fraction of cooperators when averaging over different realizations $i$, increases with time. Due to a subsequent antagonism between selection pressure towards more defectors and asymmetric exponential amplification of fluctuations during growth phase, there is only a transient increase of cooperation in a finite time window, $t\in\left[0,t_c\right]$. These findings are illustrated in a movie \footnote{See EPAPS document 1 [movie file].} showing the time evolution of the probability distribution for an ensemble of stochastic realizations.

Additional qualitative and quantitative insights can be gained from analytic calculations via a van Kampen approximation ~\cite{Kampen}, see EPAPS document~\footnote{See EPAPS Document 2 [text file]}. Thereby starting with a master equation given by Eqs.~(\ref{eq:rates}) first and higher moments of the fluctuations can be obtained. They show that  fluctuations during the first generation (i.e. doubling the initial population size on average) are by far the dominant source for the variance in the composition of the population. In addition (see below), these calculations give a strictly lower bound on the parameter regime where the cooperation time is finite and thus quantify the magnitude of fluctuations necessary to overcome the strength of selection acting against cooperators.

\begin{figure}
\centering
\includegraphics[width=0.44\textwidth]{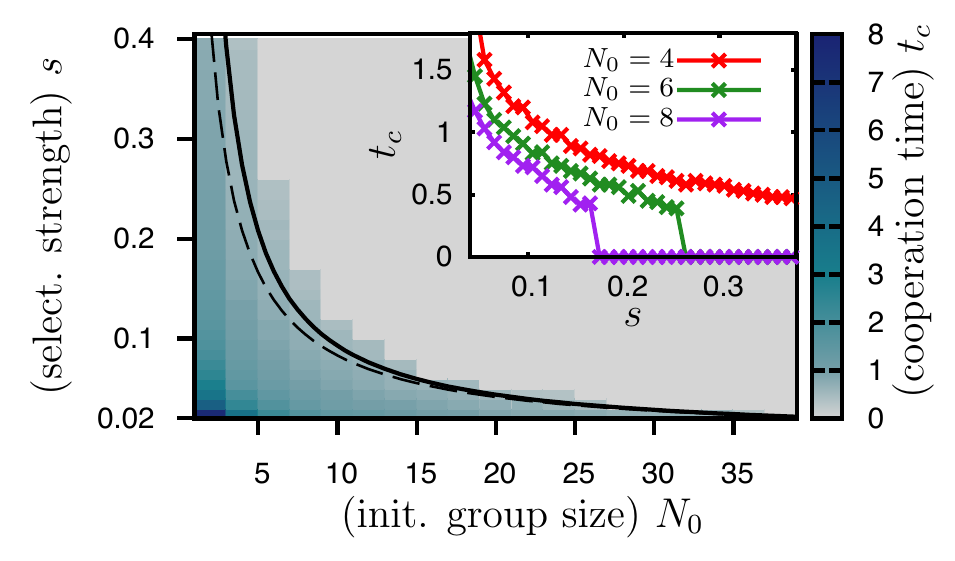}
\caption{Dependence of the cooperation time $t_c$ on the strength of selection $s$ and the initial population size $N_0$. There exist two distinct phases: the phase of transient maintained cooperation (where $t_c>0$ holds) and the phase of extinction of cooperation (where $t_c=0$). The boundary of both phases (solid line) is approximately given by $sN_0\approx p/(1+px_0)$ (dashed line). The cooperation time $t_c$ is shown for varying $s$ but fixed $N_0$ in the inset. See text and~[25]\label{Fig:phasediagram} }

\end{figure}

Fig.~\ref{Fig:phasediagram} shows the cooperation time, $t_c$, with varying selection strength, $s$, and initial population size, $N_0$. For large $s$ and $N_0$ (light grey area), $t_c$ is identical to zero, i.e. the fraction of cooperators always decreases as predicted by the deterministic  replicator dynamics, Eq.~\ref{eq:pdx}. In contrast, if $s$ and $N_0$ are sufficiently small, $t_c$ is finite. The transition between these regimes is discontinuous marked by a steep drop in the cooperation time from a finite value to zero; see Fig.~\ref{Fig:phasediagram} (inset). A strictly lower bound for the phase boundary (Fig.~\ref{Fig:phasediagram}, solid line) can be derived analytically by comparing the antagonistic effects of drift and fluctuations, see ~[25]. Its asymptotic behavior for large $N_0$ is given by $sN_0\approx p/(1+px_0)$ (Fig.~\ref{Fig:phasediagram}, dashed line). This behavior resembles the condition for neutral evolution~\cite{Kimura,Cremer2}. Indeed, for $sN_0< p/(1+px_0)$, fluctuations dominate and the system evolves neutrally. It is this neutral evolution leading to sufficiently large fluctuations which in turn - by asymmetric amplification - result in a transient increase of cooperation.

In summary, we introduced a general approach, which couples the internal evolution of a population to its growth dynamics. Both processes originate from birth and death events and are therefore naturally described by a unifying stochastic model. The standard formulations of evolutionary game theory and population dynamics emerge as special cases. Importantly, by including the coupling, our model offers the opportunity to investigate a broad range of phenomena which cannot be studied by standard approaches. We have demonstrated this for the prisoners dilemma in growing populations. Here, a transient regime of increasing cooperation can emerge by a fluctuation-induced effect. For this effect, the positive correlation between global population fitness and the level of cooperation is essential. Similar to the Luria-Delbr\"uck experiment~\cite{Delbrueck:1943p491}, initial fluctuations in the fraction of cooperators are exponentially amplified. Here, this renders it possible for cooperators to overcome the selection advantage of defectors.

In biological settings, growth is ubiquitous: populations regularly explore new habitats, or almost go extinct by external catastrophes and rebuild afterwards. For a realistic description, it is therefore necessary to relax the assumption of a decoupled population size. Especially for bacterial populations undergoing a life-cycle with a  repeated change between dispersal and maturation phases~\cite{West:2006p1249,Gore:2009p1538,Chuang01092009,Griffin}, a transient increase in cooperation may be  sufficient to overcome the dilemma of cooperation.

Financial support by the Deutsche Forschunggemeinschaft through the SFB TR12
``Symmetries and Universalities in Mesoscopic Systems" is gratefully acknowledged.

\newpage

\onecolumngrid

\setcounter{page}{1}

\begin{center}
{\bf \large Evolutionary game theory in growing populations}\\
\vspace*{0.5cm}
{\large Anna Melbinger, Jonas Cremer, and Erwin Frey}\\
\vspace*{0.5cm}
{\large Supplementary EPAPS document: conditions for the transient increase of cooperation}
\end{center}

The transient increase of cooperation emerges if initial fluctuations in the evolutionary dynamics are sufficiently large such that the asymmetrical amplification  of those can overcome the selection advantage of cheaters. In this Supplementary Material we derive the conditions for the transient increase. In particular, we give an analytical expression for the phase boundary in Fig.~2 (black line).

The full stochastic dynamics is given by the master equation determined by the birth and death rates, Eq.~(3),
\begin{align}
\frac{dP(A,B)}{dt}=&\;\;\;\;\,\Gamma_{\emptyset \to A}(A\!-\!1,B)(A\!-\!1)P(A\!-\!1,B)+\Gamma_{\emptyset \to B}(A,B\!-\!1)(B\!-\!1)P(A,B\!-\!1)\nonumber \\
&+\Gamma_{A \to \emptyset}(A\!+\!1,B)(A+1)P(A\!+\!1,B)+\Gamma_{B \to \emptyset}(A,B\!+\!1)(B\!+\!1)P(A,B+1)\nonumber \\
&- \left[\Gamma_{\emptyset \to A}(A,B)A+\Gamma_{\emptyset \to B}(A,B)B+\Gamma_{A \to \emptyset}(A,B)A+\Gamma_{B \to \emptyset}(A,B)B \right]P(A,B).\nonumber \\ \label{eq:master_equation}\tag{7}
\end{align}
Here, $A \equiv N_A$ and $B \equiv N_B$ stand for the number of individuals of both traits. We approximate the master equation upon performing a van Kampen expansion~\cite{Kampen}. To this end, we consider $A$ and $B$ as extensive variables which we write as 
\begin{align}
	A&=\Omega a(t)+\sqrt{\Omega}\xi \,\nonumber , \\ 
	B&=\Omega b(t)+\sqrt{\Omega}\mu \,.\label{eq:vanKampen}\tag{8}
\end{align} 
Here, $\Omega$ is of the order of the actual system size, and deterministically evolving densities $a(t)$ and $b(t)$ are corrected by fluctuations  $\xi$ and $\mu$. By this Ansatz the strength of fluctuations is correctly considered; their relative impact decreases like $1 / \sqrt\Omega$ with increasing system size. In the following, we consider the initial dynamics of the population when starting with a small population size $N_0$. Then, $\Omega$ is of the order $\Omega\approx N_0$. Death events can be neglected as the initial population size is far below the carrying capacity, $N_0/K\approx 0$. 

To proceed, we expand Eq.~(\ref{eq:master_equation}) in orders of $1/\sqrt{\Omega}$. The deterministic equations follow to leading order, $\mathcal{O}\left(\sqrt{\Omega}\right)$, see Eqs.~(6) with $N/K\rightarrow0$ and $x(t)=a(t)/\left[a(t)+b(t)\right]$. The next leading order, $\mathcal{O}(\Omega^0)$, results in a Fokker-Planck equation for the probability distribution of the fluctuations, $\Pi(\xi,\mu)$. The dynamics in $\Pi(\xi,\mu)$ is coupled to the deterministic equations and can be extended to include higher orders, $\mathcal{O}\left(1/\sqrt{\Omega}\right)$. From the Fokker-Planck equation for $\Pi(\xi,\mu)$, differential equations for the first moments of $\xi$ and $\mu$ can be obtained. They have the following functional form,
\begin{align}
\partial_t \langle \xi \rangle=&C_1\langle \xi \rangle +C_2 \langle \mu \rangle +\frac{1}{\sqrt{\Omega}}(C_3\langle \xi^2\rangle+C_4\langle \xi \mu \rangle +C_5\langle \mu^2\rangle) + \mathcal{O}\left(\frac{1}{\Omega}\right) , \nonumber\\
\partial_t \langle \mu \rangle=&D_1\langle \xi \rangle +D_2 \langle \mu \rangle +\frac{1}{\sqrt{\Omega}}(D_3\langle \xi^2\rangle+D_4\langle \xi \mu \rangle +D_5\langle \mu^2\rangle) + \mathcal{O}\left(\frac{1}{\Omega}\right)
\tag{9}\label{eq:firstmoment}.
\end{align}
The constants $C_i $ and $D_i$ with $i\in\{1,2,3,4,5\}$, depend on the parameters $s~,b,~c,~p$, the \emph{deterministic} parts of the composition of the population, $x(t)=a(t)/\left[a(t)+b(t)\right]$, and the population size $n(t)=a(t)+b(t)$ (in units of $\Omega$), respectively. Importantly, the second moments couple into the dynamics only through $\mathcal{O}\left(1 /  \sqrt{\Omega}\right)$ corrections. 

Neglecting these second and higher order moments, the ensuing linear equation has an unstable fixed point at $(\langle \xi \rangle, \langle \mu\rangle)^*=(0,0)$. The eigendirection with the larger (positive) eigenvalue has a component in the $\xi$-direction which is significantly larger than its component in the $\mu$-direction. As a consequence, the fluctuations in the number of cooperators ($\xi$) are amplified more strongly than those of the defectors ($\mu$); fluctuations are  asymmetrically amplified. 

Next, we consider the effect of the second moments on the dynamics. Consider a single initial state without any variance (and all other higher moments identically zero), starting the dynamics in the fixed point, $(\langle \xi \rangle, \langle \mu\rangle)^*=(0,0)$. Then, since the first moments are zero, only higher orders  in Eq.~(\ref{eq:firstmoment}) lead to deviations from the (linearly unstable) fixed point. Once such deviations are generated these are amplified by the (linearly) unstable dynamics, i.e. the first moments in Eq.~(\ref{eq:firstmoment}).
In more detail, consider the differential equations of the second moments which, for $t\rightarrow0$, have the following asymptotic form:
\begin{align}
\partial_t \langle  \xi^2\rangle=& n(1+px) \left[1+s(bx-c) \right] x\nonumber,\\
\partial_t \langle  {\xi \mu} \rangle=& 0\nonumber, \\
\partial_t \langle  \mu^2 \rangle =& n(1+px)(1+sbx)(1-x)\tag{10}.
\label{eq:2mom}
\end{align}
Starting with zero at $t=0$, both, $\langle \xi^2 \rangle$ and $\langle \mu^2 \rangle$ increase linearly in time (note that the fitness of a cooperator $1+s(bx-c) > 0$ since otherwise the birth rate would be negative). Within one generation, $t_g=1/\left[(1+px)(1+s(b-c)x )\right]$ (compare Eq. (6b)), i.e. doubling the population size on average, finite variances $\langle  \xi^2\rangle_g$ and $\langle  \mu^2\rangle_g$ are generated. This variance can be taken as a \emph{lower bound}. We even expect it to be a reasonable estimate for the actual value since the impact of the variance created in following generations on Eqs.~\ref{eq:firstmoment} is strongly suppressed by the increase in population size. 

Upon inserting the values $\langle  \xi^2\rangle_g$ and $\langle  \mu^2\rangle_g$ into Eq.~(\ref{eq:firstmoment}) one can 
now calculate the time evolution of the first moments, $\langle \xi \rangle$ and $\langle \mu \rangle$. This allows to determine the conditions necessary for a transient increase of cooperation by analyzing the fraction of cooperators $\langle\frac{A}{A+B}\rangle$; see Eqs.~(\ref{eq:vanKampen}).  The phase boundary separating the regimes of transient increase and immediate decrease of cooperation is defined by an initially stationary fraction of cooperators: $\partial_t{\langle\frac{A}{A+B}\rangle}=0$ at $t\approx 0$. 

The ensuing phase boundary is plotted in Fig.~2 (black line). The deviation from the actual (numerically determined) transition line is small for intermediate $\Omega$ and goes to zero for larger $\Omega$. By evaluating the expression in orders of $s/p$,  the lower bound of the transition line can be further simplified. To first order one finds
\begin{equation}
s=\frac{p}{n\Omega(1+px)}\tag{11},
\end{equation}
with $\Omega n=N_0$; see Fig.~2, dashed line. Note that this expression gives the asymptotically correct results for large $\Omega$. 

It is instructive to compare this result with the theory of neutral evolution~\cite{Kimura} where a condition $sN_0 \propto 1$ separates regimes of neutral and selection-dominated evolution~\cite{Kimura, Cremer2}. In the present case, for the transient increase of cooperation to occur, the system has to evolve neutrally in the initial phase to create a large enough variation in the fraction of cooperators. Then, after being asymmetriclly amplified, these fluctuations can overcome the selection pressure towards more defectors. This is mathematically reflected in Eqs.~(\ref{eq:firstmoment})  and (\ref{eq:2mom}). Initially, the second moments increase, Eqs.~(\ref{eq:2mom}), which then feeds into Eqs.~(\ref{eq:firstmoment}) and lead to an increase in the first moments. Finally, the good agreement of the phase boundary with its lower bound, reassures that the variation in cooperators fraction is mainly generated at the beginning of the dynamics.


\begin{thebibliography}{22}
\expandafter\ifx\csname natexlab\endcsname\relax\def\natexlab#1{#1}\fi
\expandafter\ifx\csname bibnamefont\endcsname\relax
  \def\bibnamefont#1{#1}\fi
\expandafter\ifx\csname bibfnamefont\endcsname\relax
  \def\bibfnamefont#1{#1}\fi
\expandafter\ifx\csname citenamefont\endcsname\relax
  \def\citenamefont#1{#1}\fi
\expandafter\ifx\csname url\endcsname\relax
  \def\url#1{\texttt{#1}}\fi
\expandafter\ifx\csname urlprefix\endcsname\relax\def\urlprefix{URL }\fi
\providecommand{\bibinfo}[2]{#2}
\providecommand{\eprint}[2][]{\url{#2}}

\bibitem[{\citenamefont{Ewens}(2004)}]{Ewens}
\bibinfo{author}{\bibfnamefont{W.~J.} \bibnamefont{Ewens}},
  \emph{\bibinfo{title}{Mathematical Population Genetics}}
  (\bibinfo{publisher}{Springer}, \bibinfo{address}{New York},
  \bibinfo{year}{2004}).

\bibitem[{\citenamefont{{Maynard Smith}}(1982)}]{Maynard}
\bibinfo{author}{\bibfnamefont{J.}~\bibnamefont{{Maynard Smith}}},
  \emph{\bibinfo{title}{Evolution and the Theory of Games}}
  (\bibinfo{publisher}{Cambridge University Press},
  \bibinfo{address}{Cambridge}, \bibinfo{year}{1982}).

\bibitem[{\citenamefont{Blythe and McKane}(2007)}]{Blythe:2007}
\bibinfo{author}{\bibfnamefont{R.~A.} \bibnamefont{Blythe}} \bibnamefont{and}
  \bibinfo{author}{\bibfnamefont{A.~J.} \bibnamefont{McKane}},
  \bibinfo{journal}{J. Stat. Mech.} \textbf{\bibinfo{volume}{2007}},
  \bibinfo{pages}{P07018} (\bibinfo{year}{2007}).

\bibitem[{\citenamefont{Murray}(2002)}]{Murray}
\bibinfo{author}{\bibfnamefont{J.}~\bibnamefont{Murray}},
  \emph{\bibinfo{title}{Mathematical Biology}}, 1
  (\bibinfo{publisher}{Springer}, \bibinfo{year}{2002}).

\bibitem[{\citenamefont{Hastings}(1997)}]{Hastings1997}
\bibinfo{author}{\bibfnamefont{A.}~\bibnamefont{Hastings}},
  \emph{\bibinfo{title}{Population Biology: Concepts and Models}}
  (\bibinfo{publisher}{Springer}, \bibinfo{year}{1997}).

\bibitem[{\citenamefont{Hibbing et~al.}(2010)\citenamefont{Hibbing, Fuqua,
  Parsek, and Peterson}}]{Hibbing2010}
\bibinfo{author}{\bibfnamefont{M.~E.} \bibnamefont{Hibbing}},
  \bibinfo{author}{\bibfnamefont{C.}~\bibnamefont{Fuqua}},
  \bibinfo{author}{\bibfnamefont{M.~R.} \bibnamefont{Parsek}},
  \bibnamefont{and} \bibinfo{author}{\bibfnamefont{S.~B.}
  \bibnamefont{Peterson}}, \bibinfo{journal}{Nat. Rev. Microbiol.}
  \textbf{\bibinfo{volume}{8}}, \bibinfo{pages}{15} (\bibinfo{year}{2010}).

\bibitem[{\citenamefont{Roughgarden}(1971)}]{Roughgarden}
\bibinfo{author}{\bibfnamefont{J.}~\bibnamefont{Roughgarden}},
  \bibinfo{journal}{{Ecology}} \textbf{\bibinfo{volume}{52}},
  \bibinfo{pages}{453} (\bibinfo{year}{1971}).

\bibitem[{\citenamefont{West et~al.}(2006)\citenamefont{West, Griffin, Gardner,
  and Diggle}}]{West:2006p1249}
\bibinfo{author}{\bibfnamefont{S.~A.} \bibnamefont{West}},
  \bibinfo{author}{\bibfnamefont{A.~S.} \bibnamefont{Griffin}},
  \bibinfo{author}{\bibfnamefont{A.}~\bibnamefont{Gardner}}, \bibnamefont{and}
  \bibinfo{author}{\bibfnamefont{S.~P.} \bibnamefont{Diggle}},
  \bibinfo{journal}{Nat. Rev. Microbiol.} \textbf{\bibinfo{volume}{4}},
  \bibinfo{pages}{597} (\bibinfo{year}{2006}).

\bibitem[{\citenamefont{Hauert et~al.}(2006)\citenamefont{Hauert, Holmes, and
  Doebeli}}]{Hauert3}
\bibinfo{author}{\bibfnamefont{C.}~\bibnamefont{Hauert}},
  \bibinfo{author}{\bibfnamefont{M.}~\bibnamefont{Holmes}}, \bibnamefont{and}
  \bibinfo{author}{\bibfnamefont{M.}~\bibnamefont{Doebeli}},
  \bibinfo{journal}{Proc. R. Soc. Lond. B.} \textbf{\bibinfo{volume}{273}},
  \bibinfo{pages}{2565} (\bibinfo{year}{2006}).

\bibitem[{\citenamefont{Cressman and Vickers}(1997)}]{Cressman}
\bibinfo{author}{\bibfnamefont{R.}~\bibnamefont{Cressman}} \bibnamefont{and}
  \bibinfo{author}{\bibfnamefont{G.}~\bibnamefont{Vickers}},
  \bibinfo{journal}{J. Theor. Biol.} \textbf{\bibinfo{volume}{184}},
  \bibinfo{pages}{359} (\bibinfo{year}{1997}).

\bibitem[{\citenamefont{Kimura}(1983)}]{Kimura}
\bibinfo{author}{\bibfnamefont{M.}~\bibnamefont{Kimura}},
  \emph{\bibinfo{title}{The Neutral Theory of Molecular Evolution}}
  (\bibinfo{publisher}{Cambridge University Press},
  \bibinfo{address}{Cambridge}, \bibinfo{year}{1983}).

\bibitem[{\citenamefont{Nowak et~al.}(2004)\citenamefont{Nowak, Sasaki, Taylor,
  and Fudenberg}}]{nowak-2004-428}
\bibinfo{author}{\bibfnamefont{M.~A.} \bibnamefont{Nowak}},
  \bibinfo{author}{\bibfnamefont{A.}~\bibnamefont{Sasaki}},
  \bibinfo{author}{\bibfnamefont{C.}~\bibnamefont{Taylor}}, \bibnamefont{and}
  \bibinfo{author}{\bibfnamefont{D.}~\bibnamefont{Fudenberg}},
  \bibinfo{journal}{Nature} \textbf{\bibinfo{volume}{428}}, \bibinfo{pages}{646
  } (\bibinfo{year}{2004}).

\bibitem[{\citenamefont{Traulsen et~al.}(2005)\citenamefont{Traulsen, Claussen,
  and Hauert}}]{Traulsen}
\bibinfo{author}{\bibfnamefont{A.}~\bibnamefont{Traulsen}},
  \bibinfo{author}{\bibfnamefont{J.~C.} \bibnamefont{Claussen}},
  \bibnamefont{and} \bibinfo{author}{\bibfnamefont{C.}~\bibnamefont{Hauert}},
  \bibinfo{journal}{Phys. Rev. Lett.} \textbf{\bibinfo{volume}{95}},
  \bibinfo{pages}{238701} (\bibinfo{year}{2005}).

\bibitem[{\citenamefont{Cremer et~al.}(2009)\citenamefont{Cremer, Reichenbach,
  and Frey}}]{Cremer2}
\bibinfo{author}{\bibfnamefont{J.}~\bibnamefont{Cremer}},
  \bibinfo{author}{\bibfnamefont{T.}~\bibnamefont{Reichenbach}},
  \bibnamefont{and} \bibinfo{author}{\bibfnamefont{E.}~\bibnamefont{Frey}},
  \bibinfo{journal}{New J. Phys.} \textbf{\bibinfo{volume}{11}},
  \bibinfo{pages}{093029} (\bibinfo{year}{2009}).

\bibitem[{\citenamefont{Moran}(1964)}]{Moran}
\bibinfo{author}{\bibfnamefont{P.~A.} \bibnamefont{Moran}},
  \emph{\bibinfo{title}{The {S}tatistical {P}rocesses of {E}volutionary
  {T}heory}} (\bibinfo{publisher}{Clarendon Press Oxford},
  \bibinfo{address}{Oxford}, \bibinfo{year}{1964}).

\bibitem[{\citenamefont{Nowak}(2006)}]{NowakCooperation}
\bibinfo{author}{\bibfnamefont{M.~A.} \bibnamefont{Nowak}},
  \bibinfo{journal}{Science} \textbf{\bibinfo{volume}{314}},
  \bibinfo{pages}{1560} (\bibinfo{year}{2006}).

\bibitem[{\citenamefont{Gore et~al.}(2009)\citenamefont{Gore, Youk, and van
  Oudenaarden}}]{Gore:2009p1538}
\bibinfo{author}{\bibfnamefont{J.}~\bibnamefont{Gore}},
  \bibinfo{author}{\bibfnamefont{H.}~\bibnamefont{Youk}}, \bibnamefont{and}
  \bibinfo{author}{\bibfnamefont{A.}~\bibnamefont{van Oudenaarden}},
  \bibinfo{journal}{Nature} \textbf{\bibinfo{volume}{459}},
  \bibinfo{pages}{253}   (\bibinfo{year}{2009}).

\bibitem[{\citenamefont{Chuang et~al.}(2009)\citenamefont{Chuang, Rivoire, and
  Leibler}}]{Chuang01092009}
\bibinfo{author}{\bibfnamefont{J.~S.} \bibnamefont{Chuang}},
  \bibinfo{author}{\bibfnamefont{O.}~\bibnamefont{Rivoire}}, \bibnamefont{and}
  \bibinfo{author}{\bibfnamefont{S.}~\bibnamefont{Leibler}},
  \bibinfo{journal}{Science} \textbf{\bibinfo{volume}{323}},
  \bibinfo{pages}{272} (\bibinfo{year}{2009}).

\bibitem[{\citenamefont{Griffin et~al.}(2004)\citenamefont{Griffin, West, and
  Buckling}}]{Griffin}
\bibinfo{author}{\bibfnamefont{A.~S.} \bibnamefont{Griffin}},
  \bibinfo{author}{\bibfnamefont{S.~A.} \bibnamefont{West}}, \bibnamefont{and}
  \bibinfo{author}{\bibfnamefont{A.}~\bibnamefont{Buckling}},
  \bibinfo{journal}{Nature} \textbf{\bibinfo{volume}{430}},
  \bibinfo{pages}{1024} (\bibinfo{year}{2004}).

\bibitem[{\citenamefont{Verhulst}(1838)}]{Verhulst}
\bibinfo{author}{\bibfnamefont{P.~F.} \bibnamefont{Verhulst}},
  \bibinfo{journal}{Corresp. Math. Phys.} \textbf{\bibinfo{volume}{10}},
  \bibinfo{pages}{113} (\bibinfo{year}{1838}).

\bibitem[{\citenamefont{Kampen}(2001)}]{Kampen}
\bibinfo{author}{\bibfnamefont{N.~V.} \bibnamefont{Kampen}},
  \emph{\bibinfo{title}{Stochastic Processes in Physics and Chemistry}} (\bibinfo{publisher}{Elsevier},
   \bibinfo{address}{Amsterdam}, \bibinfo{year}{2007}), \bibinfo{edition}{3rd} ed.

\bibitem[{\citenamefont{Luria and Delbr\"{u}ck}(1943)}]{Delbrueck:1943p491}
\bibinfo{author}{\bibfnamefont{S.~E.} \bibnamefont{Luria}} \bibnamefont{and}
  \bibinfo{author}{\bibfnamefont{M.}~\bibnamefont{Delbr\"{u}ck}},
  \bibinfo{journal}{Genetics} \textbf{\bibinfo{volume}{28}},
  \bibinfo{pages}{491} (\bibinfo{year}{1943}).

\end{thebibliography}

\begin{thebibliography}{1}

\bibitem{Kampen}
N.G.~Van Kampen.
\newblock {\em Stochastic Processes in Physics and Chetry (North-Holland
  Personal Library)}.
\newblock North Holland, 2nd edition, 2001.

\bibitem{Kimura}
M.~Kimura.
\newblock {\em The Neutral Theory of Molecular Evolution}.
\newblock Cambridge University Press, Cambridge, 1983.

\bibitem{Cremer2}
J.~Cremer, T.~Reichenbach, and E.~Frey.
\newblock The edge of neutral evolution in social dilemmas.
\newblock {\em New J. Phys.}, 11:093029, 2009.

\end{thebibliography}
\end{document}